\documentclass{PoS}
\usepackage{epsfig}

\newcommand{\kmps}{km s\ensuremath{^{-1}}}
\newcommand{\Msun}{M\ensuremath{_\odot}}

\title{Transport of charged dust grains into the galactic halo}

\ShortTitle{Transport of charged dust grains into the galactic halo}

\author{\speaker{Sergey Khoperskov} \thanks{RFBR grants 14-02-00604, 14-02-31456}\\
        Institute of Astronomy of the Russian Academy of Sciences, Pyatnitskaya st., 48, 119017 Moscow, Russia \\ Sternberg Astronomical Institute, Moscow MV Lomonosov State University, Universitetskij pr., 13, 119992 Moscow, Russia\\
        E-mail: \email{khoperskov@inasan.ru}}

\author{Yuri Shchekinov\thanks{RFBR grant 12-02-00917} \\
        Southern Federal University, Sorge Str. 5, Rostov-on-Don 344090, Russia\\
        E-mail: \email{yus@sfedu.ru}}

\abstract{We develop a 3D dynamical model of dust outflows from galactic discs. The outflows are initiated by multiple SN 
explosions in a magnetized interstellar medium (ISM) with a gravitationally stratified density distribution. Dust grains are treated as particles 
in cells interacting collisionally with gas, and forced by stellar radiation of the disc and Lorenz force.  We show that magnetic 
field plays a crucial role in accelerating the charged dust grains and expelling them out of the disc: in 10--20~Myr they can be 
elevated at distances up to 10~kpc above the galactic plane. The dust-to-gas ratio in the outflowing medium varies in the range 
$5 \cdot 10^{-4} - 5 \cdot 10^{-2}$ along the vertical stream. Overall the dust mass loss rate depends on 
the parameters of ISM and may reach up to $3\times 10^{-2}$~\Msun~yr$^{-1}$.}

\FullConference{The Life Cycle of Dust in the Universe: Observations, Theory, and Laboratory Experiments - LCDU 2013,\\
		18-22 November 2013\\
		Taipei, Taiwan}

\begin{document}

\section{Introduction}
\vskip -0.021\vsize The dominant sources of dust in the Milky Way are the AGB stars~\cite{2009ApJ...698.1136G}, supernovae and young stellar objects 
and cool dense ISM regions~\cite{1998ApJ...501..643D}. 
Stellar winds and jets from YSO supply dust into the nearest interstellar medium where it is supposed to mix with ambient gas. 
SNe ejecta are also thought to be an efficient source of dust, though it is unclear what fraction of dust survives in the reverse shocks 
of SNe remnants~\cite{2012ApJ...748...12S}. However, it is absolutely clear that the sources of dust concentrate basically within 
the galactic stellar disc. 

At the same time, there are copious evidences of dust extending in the vertical direction up to tens of the scale height of the stellar 
thin disc~\cite{2006A&A...445..123I,2007A&A...471L...1K,1997AJ....114.2463H,2009AJ....137.3000H}. 
In the radial direction dust may occupy at least twice as large a disc as the stellar one. In addition, more recently evidences of dust 
present in the intergalactic medium have also appeared in the literature~\cite{2010MNRAS.405.1025M}. 

\begin{figure}[!b]
\begin{center}
\includegraphics[width=0.49\hsize]{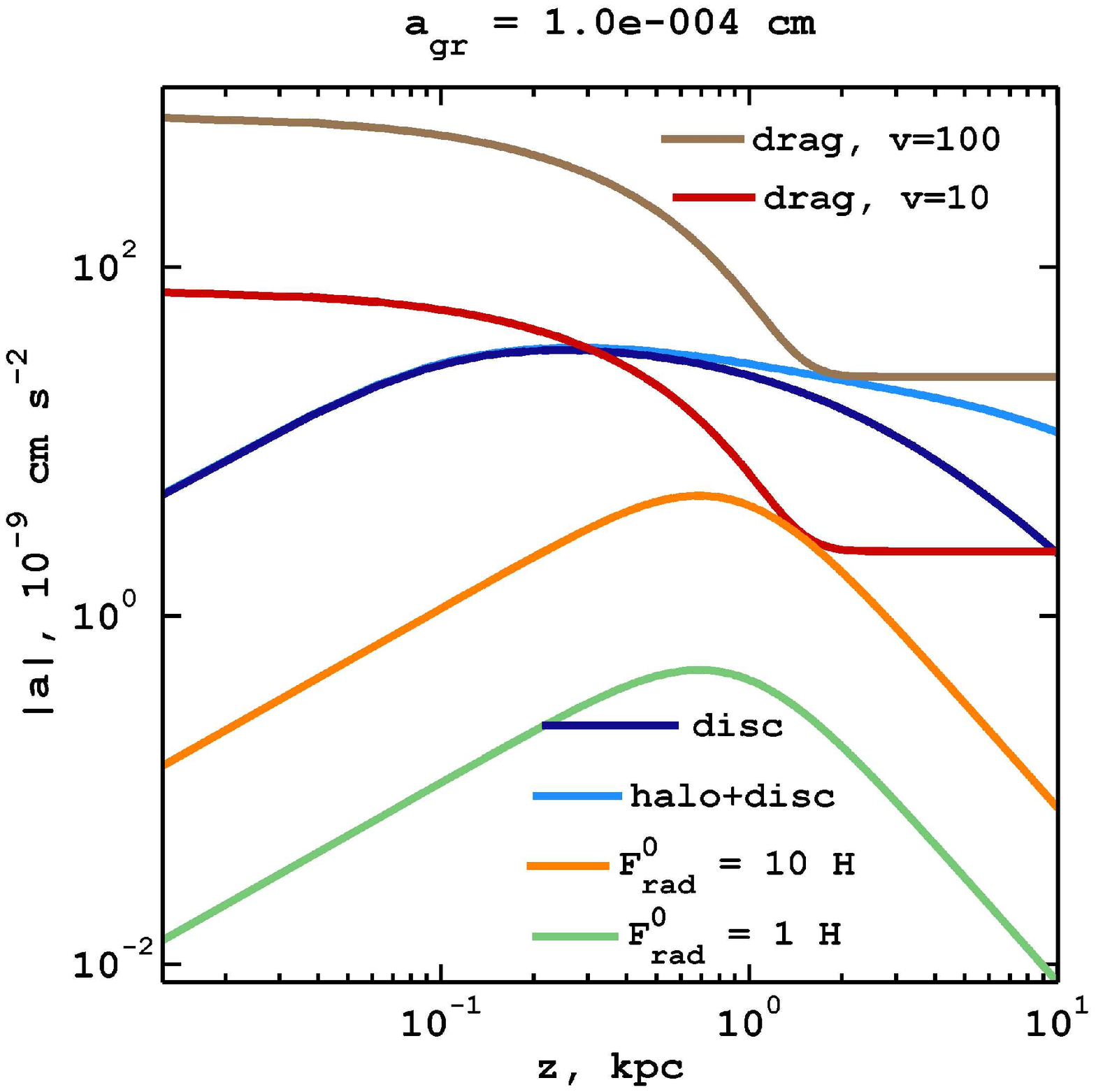}
\includegraphics[width=0.49\hsize]{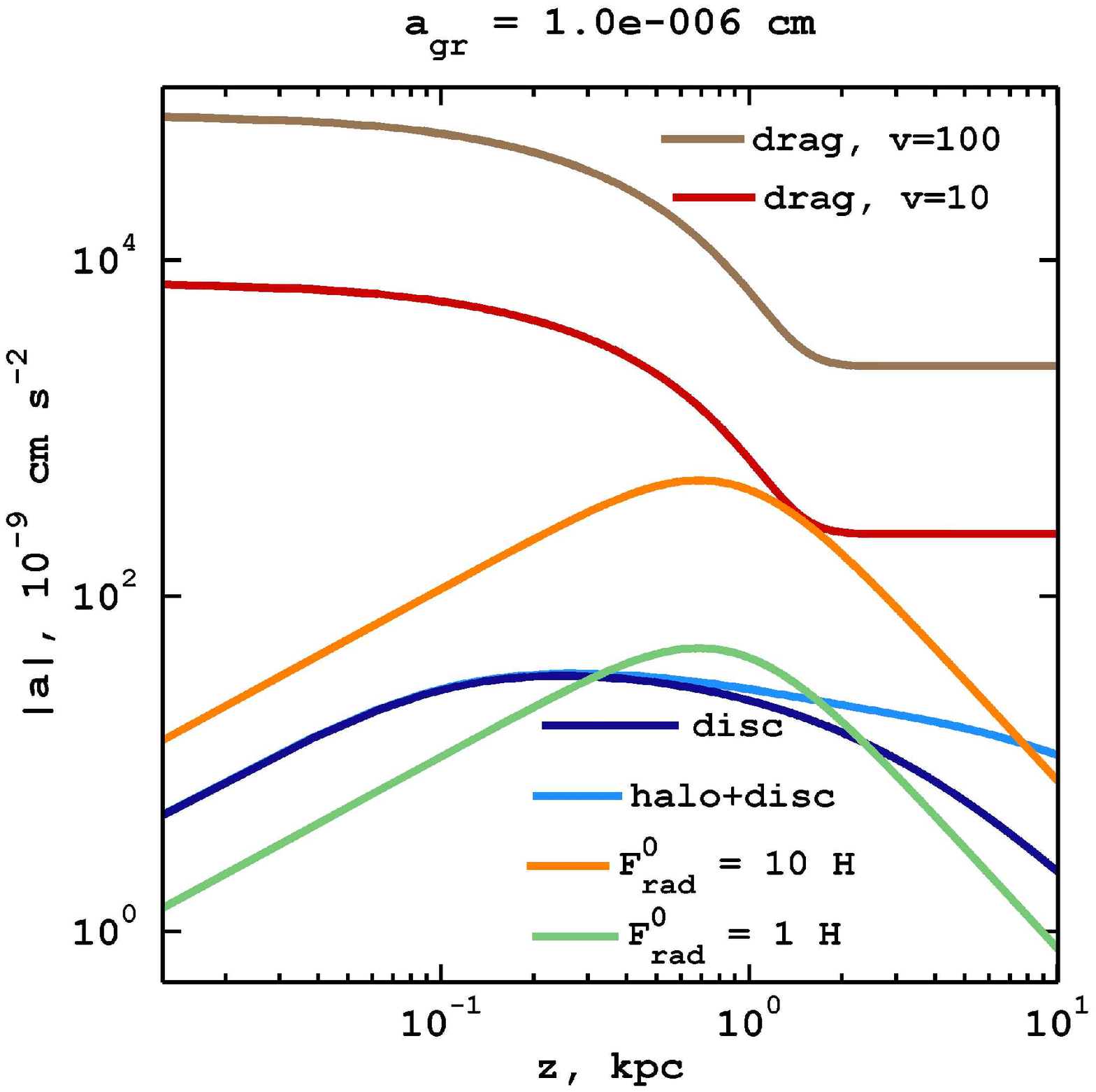}
\end{center}
\caption{Contributions to the vertical acceleration of dust grains of radius 
$a=10^{-4}$~cm (left panel) and $a=10^{-6}$~cm~(right panel): 
gravity from the disc (dark blue), from the disc and the halo~(light blue); radiation pressure for $1$ (green) and $10$~Habing 
fluxes (orange); drag forces for a grain moving through the gaseous halo with a velocity $v = 10$~\kmps~(red) and $100$~\kmps~(brown).}\label{fig::accel}
\end{figure}

In this contribution we briefly describe our results of simulations of dust transport within a 3D N-body/hydrodynamical framework 
on scales of the galactic halo. Aiming to understand of how the galactic halo and galaxy outskirts are enriched with dust we 
developed a numerical  
scenario of dust driven by a combined action of stellar radiation pressure, multiple supernova explosions and Lorenz forces  
in a magnetized gravitationally stratified ISM.

\section{Model}
\vskip -0.021\vsize We consider a dust-gaseous mixture in a 20~kpc height column based on a $1{\rm~ kpc}\times 1{\rm~kpc}$ square in the plane immersed into 
the gravitation field of the stellar disc and the dark matter halo. Description of gas dynamics is based on the TVD~MUSCL scheme with a
$10$~pc spatial resolution. Dust grains are treated within a collisionless $N$-body ($10^7$~probe particles) description. Such 
an approach allows us to treat separately motions of dust and gas in the presence of magnetic and radiation fields, mutual interaction 
of dust and gas components is due to the collisional friction: see Figure~\ref{fig::accel}.

Initially the gaseous disc is in the hydrostatic equilibrium, with a $\beta=1$ magnetic field parallel to the disc plane; dust grains 
occupy the 2~kpc layer ($|z|<1$~kpc) with the standard dust-to-gas ratio $0.01$. SN  energy $E_{51} =E/10^{51}$~erg is injected into the 
disc with the scale height $h_{\rm sn} \approx 100$~pc and with the current rate for the Milky Way disc 
of $(370)^{-1}$~yr$^{-1}$ (type I + type II). Thus, during a $10^7$~yr period we randomly distribute 
$\approx 2 \cdot 10^3$~SN explosions within the computational domain. 
Each SNe is represented by energy $E_{51}=1$ injected into the computational cell instantaneously. Our description of gaseous component is similar to the one developed by de Avillez and Breitschwerdt for modelling a realistic ISM heated by multiple SNe explosions~\cite{2005A&A...436..585D}.

\begin{figure}[!t]
\includegraphics[height=0.26\vsize]{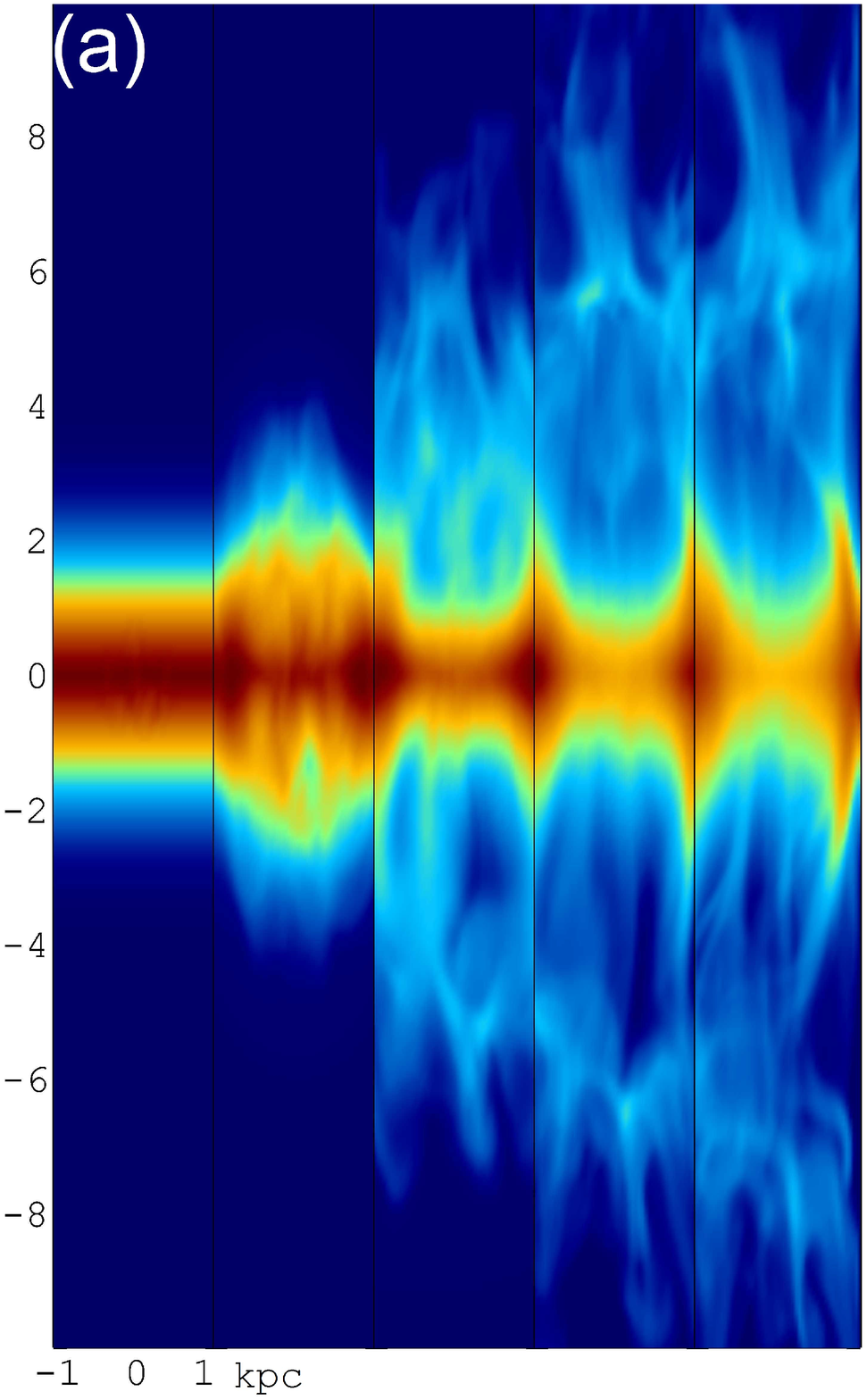}
\includegraphics[height=0.26\vsize]{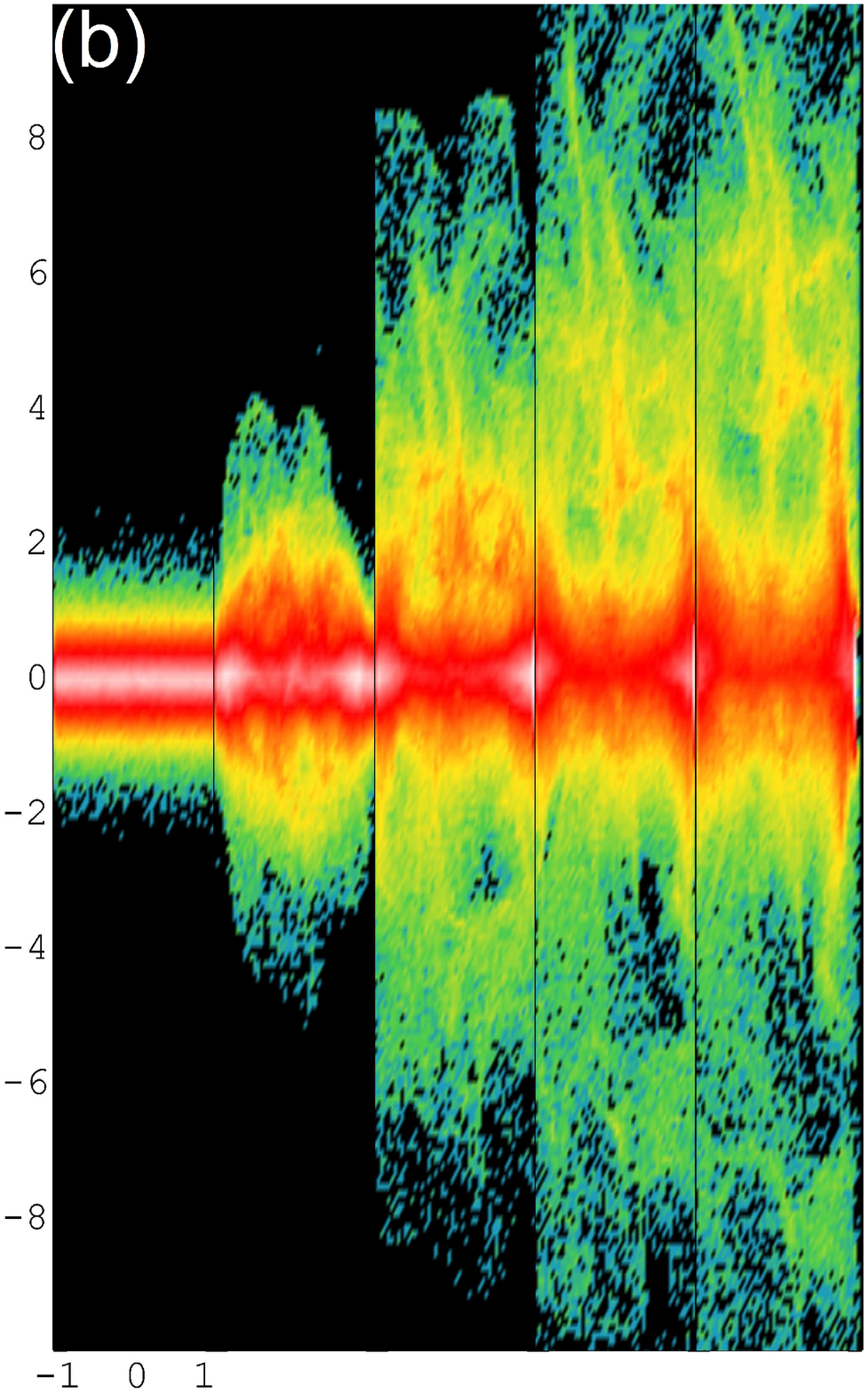}
\includegraphics[height=0.26\vsize]{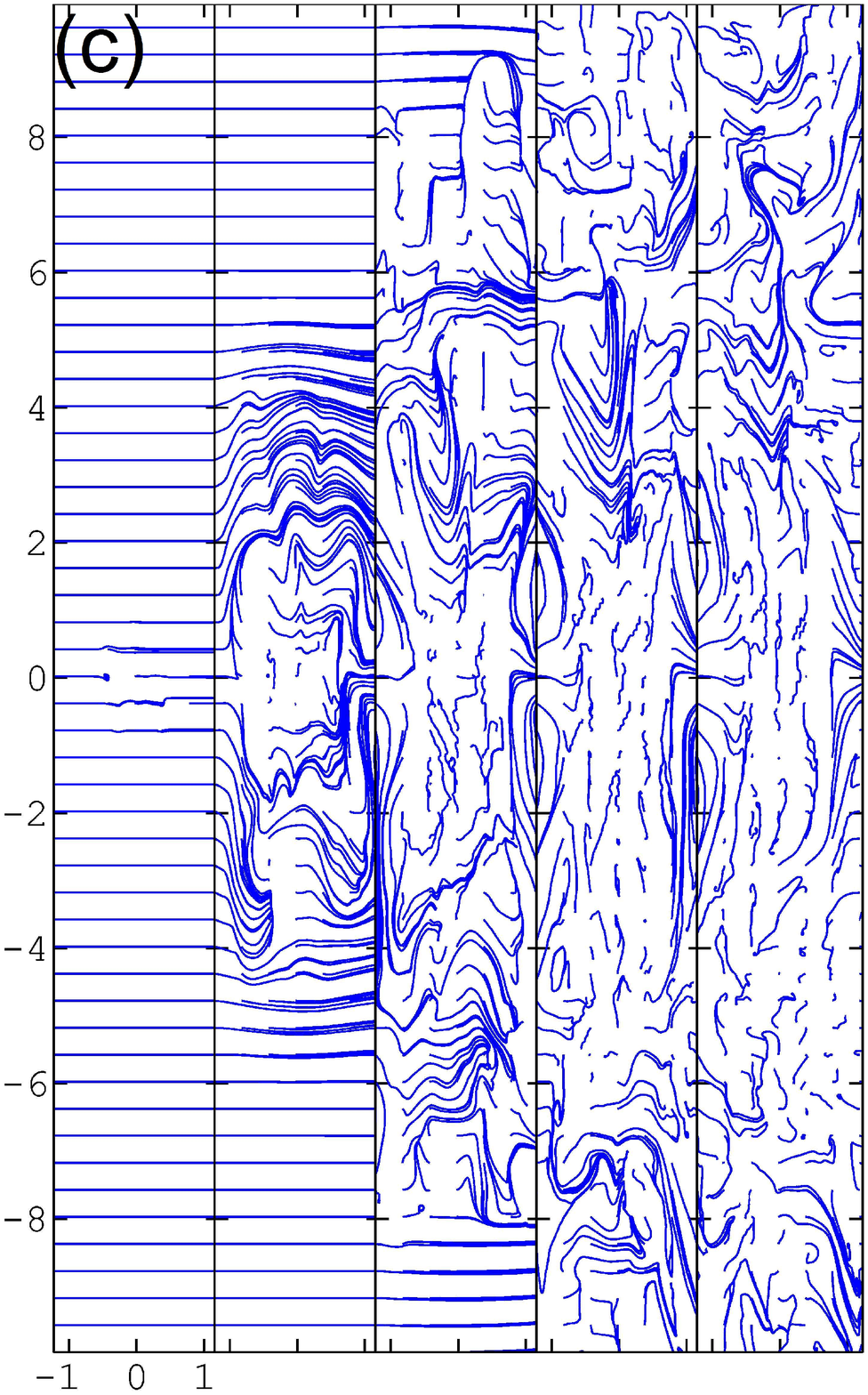}
\includegraphics[height=0.26\vsize]{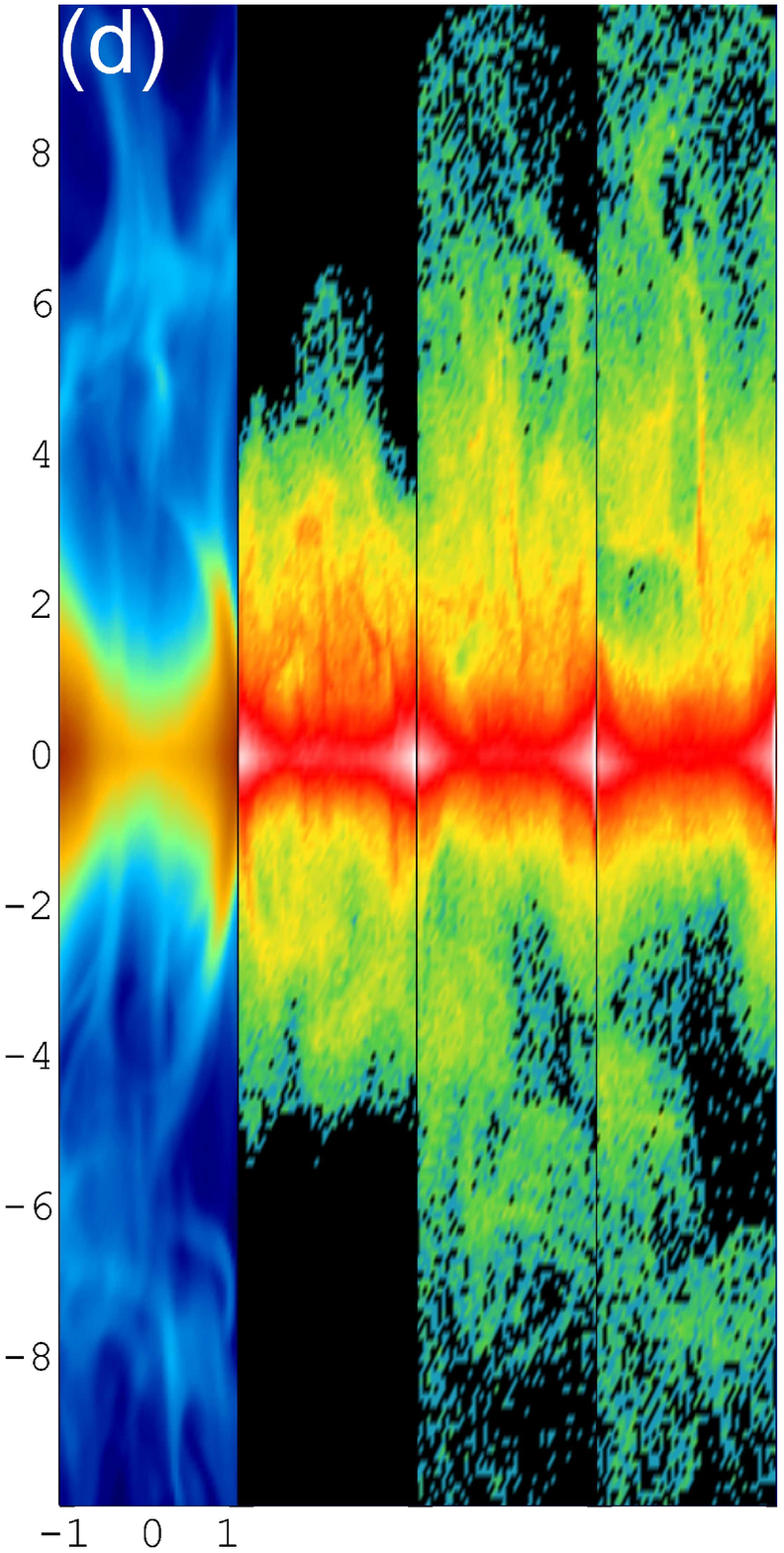}
\caption{Evolution of gas column density~(a),  dust column density~(b) and magnetic field structure~(c) at $t = 2; 7; 11; 15$~Myr. 
Panel~(d) is the distribution of gas (left) and dust grains of different sizes ($a=10^{-4}$; $10^{-5}$; $10^{-6}$~cm --- from left to right) at $t=15$~Myr.}\label{fig::evolution}
\end{figure}

\section{Results}

\begin{itemize}
\item{} Supernovae explosions along with radiation pressure from associated stellar population with a flux of $1-10$~Habing 
drive dust outflows whose overall dynamics depends on gas density and plasma-$\beta$ in the plane. Gas is expelled only by the shock waves from SNe.
\item{} An asymmetry between positive and negative altitudes is clearly seen: Fig.~\ref{fig::evolution}~(a,b,c). This is due to 
a particular choice of the random spatial distribution of SNe~explosions. 
\item{} Due to a predominance of gravity $\propto a^3$ over collisional coupling and radiation force $\propto a^2$ larger 
particles are elevated slower than the smaller ones, as seen from Fig.~\ref{fig::evolution}~(d). 
\item{} Even though initially dust and gas are assumed to be well mixed, during their elevation they separate such that 
the dust-to-gas ratio varies 
from point to point: Fig.~\ref{fig::34}~(left). 
\item{} There are two regimes of dust outflows: 1) fast motion with the rate $10^{-3} - 3\cdot 10^{-2}$~\Msun yr$^{-1}$, 
and 2) nearly constant dust mass loss rate $\sim 4\cdot 10^{-5} - 3\cdot 10^{-3}$~\Msun~yr$^{-1}$: Figure~\ref{fig::34}~(right). The dust mass loss rate varies from 
model to model, however the two regimes are a common feature.
\item{} Under the action of SNe shocks magnetic lines bend with a considerable growth of the vertical component  
in a few dynamical times. It works as an additional efficient mechanism accelerating charged dust particles upwards. 
\end{itemize} 

\begin{figure}[!t]
\includegraphics[width=0.49\hsize]{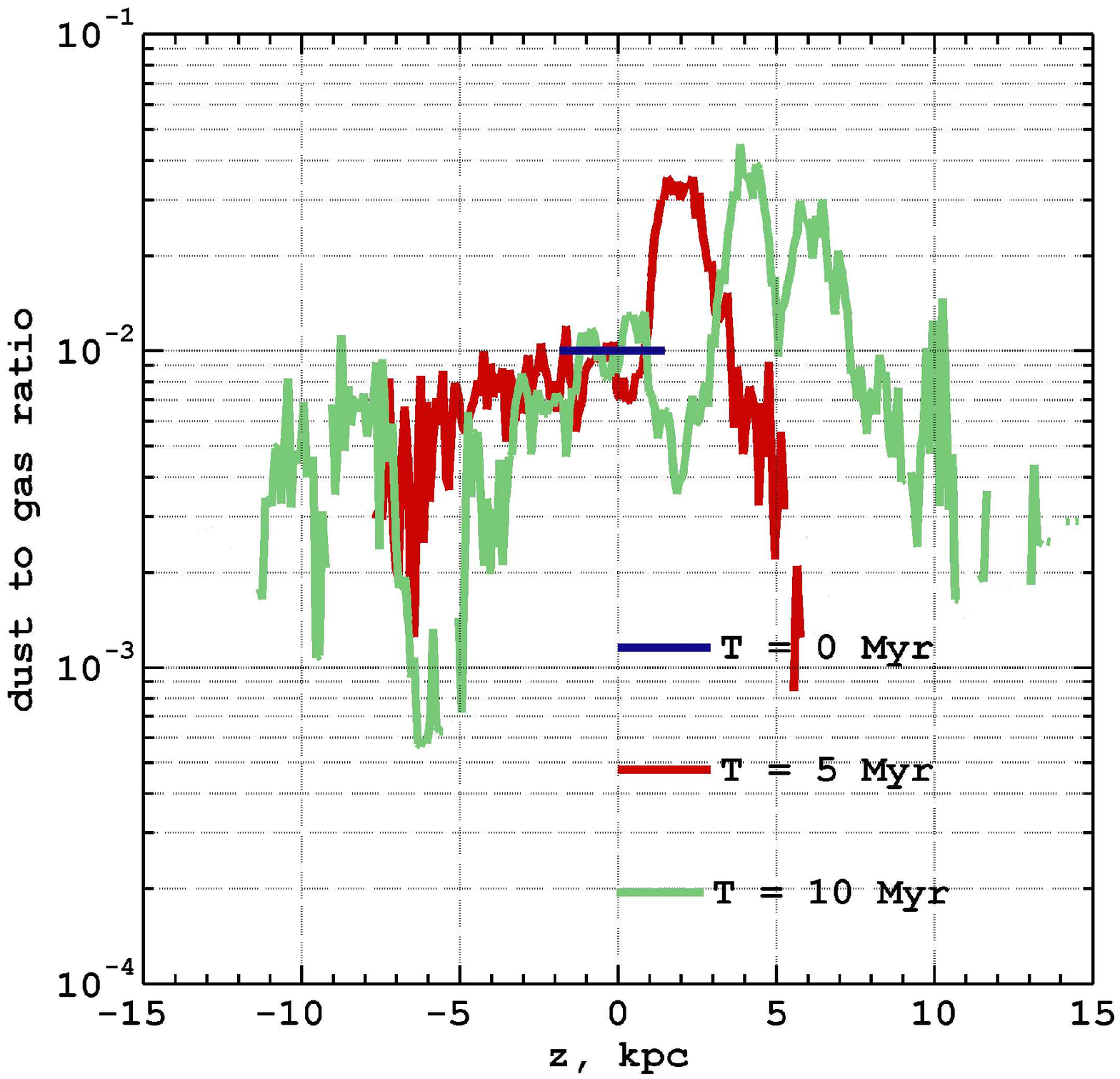}
\includegraphics[width=0.51\hsize]{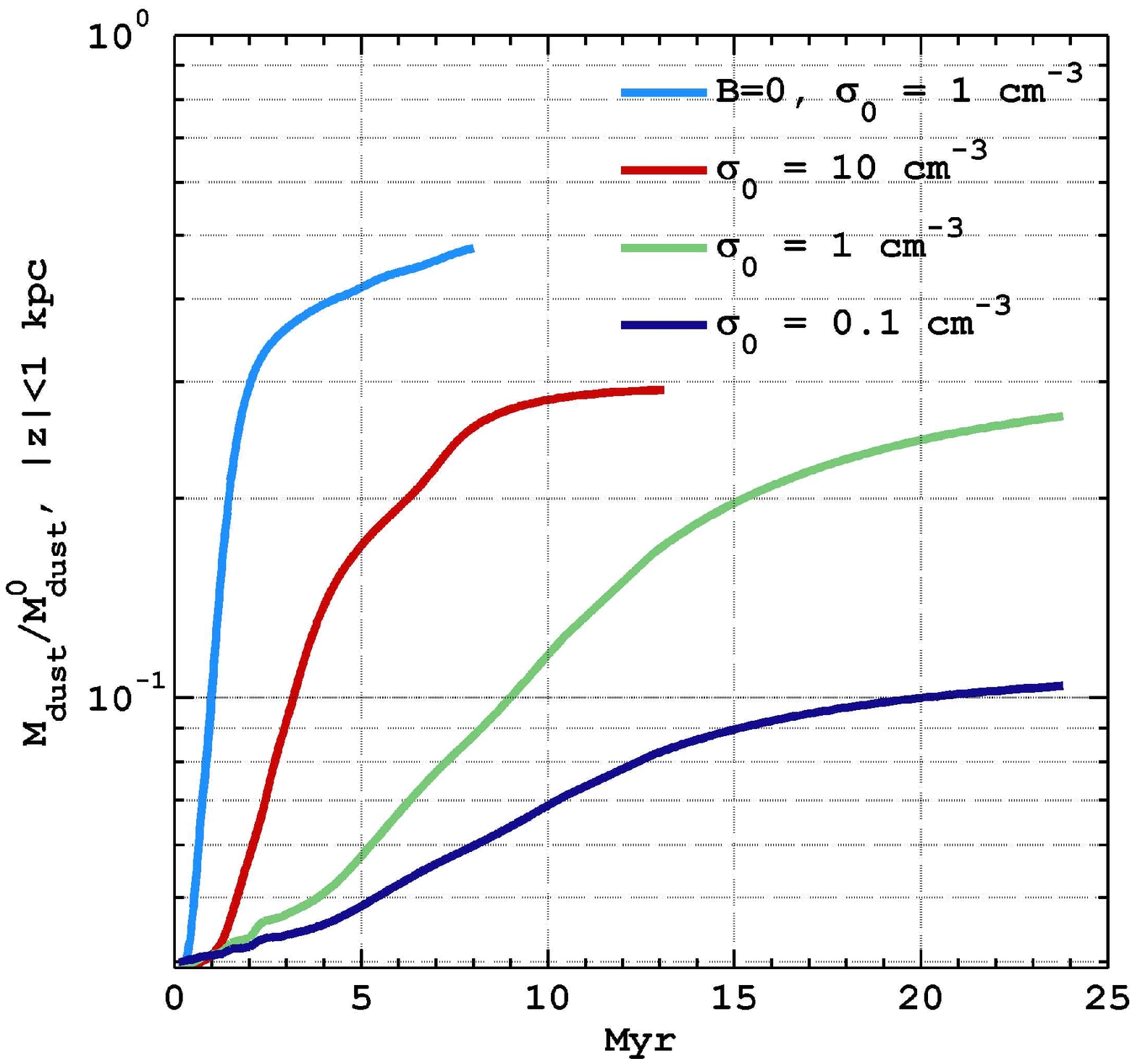}
\caption{Left: gas-to-dust ratio along the vertical direction at $t = 0$~(blue line), $t = 5$~(red line) and $10$~(green)~Myr. Right: fraction of dust mass above 1~kpc: light blue --- $\beta = 0$ and $\sigma_0 = 1$~cm$^{-3}$; 
in other simulations $\beta = 1$: red --- $\sigma_0 = 10$~cm$^{-3}$, green --- $\sigma_0 = 1$~cm$^{-3}$ and 
dark blue --- $\sigma_0 = 0.1$~cm$^{-3}$.}
\end{figure}\label{fig::34}


\section{Summary}

\vskip -0.021\vsize Multiple supernovae explosions in the galactic thin stellar disc drive gas and dust into the halo. A large amount of dust can be 
transported upto $10$~kpc in $10-20$~Myr. Ram pressure from supernovae explosions reorganizes magnetic field structure such that it 
efficiently accelerates dust grains. The dust outflow rate strongly depends on the plasma-$\beta$ parameter and 
gas density in the disc. 
The estimated dust mass loss rate is in the range from $10^{-3}$~\Msun~yr$^{-1}$ up to $3\times 10^{-2}$~\Msun~yr$^{-1}$. Such a high
rate is kept though on short time scales $2-10$~Myr.

\end{document}